\documentclass[11pt, reqno]{amsart}
\usepackage{booktabs}
\usepackage{csquotes}
\usepackage[english]{babel}
\usepackage{graphics,amsmath,amssymb,epsfig,stmaryrd,color}
\usepackage{subfigure,amsfonts,geometry,array}
\usepackage{graphicx}
\usepackage{amsthm}
\usepackage{flafter}

\usepackage{hyperref}

\usepackage[
backend=biber,
style=apa,
sorting=nyt
]{biblatex}
\DeclareSourcemap{
  \maps[datatype=bibtex]{
    \map{
      \step[fieldset=issn, null]
      \step[fieldset=doi, null]
      \step[fieldset=url, null]
      \step[fieldset=urldate, null]
    }
  }
}

\hypersetup{
    colorlinks=true,
    urlcolor=blue,
    linkcolor=blue,
    citecolor=blue
}

\def\esp{\mathbf{E}}

\def\bk{\bigskip

\noindent}
\newtheorem{defi}{Definition}
\newtheorem{ex}{Example}
\newtheorem{theoreme}{Theorem}

\newtheorem{proposition}{Proposition}

\begin{document}
\title{Extreme dependence for multivariate data}
\author{Damien Bosc$^\dag$ and  Alfred Galichon$^\ddag$}
\date{First version is March 19, 2010. The present version is of April 6, 2013.

$^\dag$Corresponding author. EDF R\&D, 1 avenue du Général de Gaulle, 92140, Clamart, France. Email: \url{damien.bosc@polytechnique.edu}. Bosc acknowledges the support of the AXA Research Fund, AXA Investment Managers and the Investment Solutions Quantitative Team.

$^\ddag$Sciences Po, Department of Economics, 28 rue des Saints-Pères, 75007 Paris, France. Galichon's research in this area has been supported by FiME, Laboratoire de Finance des March\'{e}s de l'Energie (\url{www.fime-lab.org}).

Appeared in \textit{Quantitative Finance}, Volume 14, Issue 7, March 2014, \url{https://doi.org/10.1080/14697688.2014.886777}.
}
\maketitle
\setcounter{page}{1}

\begin{abstract}
This article proposes a generalized notion of extreme multivariate dependence between two random vectors which relies on the extremality of the cross-covariance matrix between these two vectors. Using a partial ordering on the cross-covariance matrices, we also generalize the notion of positive upper
dependence. We then proposes a means to quantify the strength of the dependence between two given multivariate series and to increase this strength while preserving the marginal distributions. This allows for the design of stress-tests of the dependence between two sets of financial variables, that can be useful in portfolio management or derivatives pricing.\\


\end{abstract}

\section{Introduction}

Extreme dependence, and the closely related notion of comonotonicity are
important concepts in various fields. It is central in the economics of
insurance (following the seminal work of \cite{Borch}, \cite{Arrow63}, and \cite{Arrow70}), in economic theory (see \cite{Yaari}, \cite{LM}, and \cite{Schmeidler}), in statistics (see \cite{dallaglio}, \cite{Ru91}%
, \cite{rachev}, \cite{zolotarev}) as well as in financial risk management (see the recent book by \cite{MS} and references therein).\bigskip

\noindent The notion of extreme (positive) dependence or comonotonicity for univariate random variables goes back to the work of \cite{Hoeffding} and \cite{Frechet}. Two real random variables $%
(X,Y)$  are comonotonic if their cumulative distribution
function satisfy $F_{X,Y}(x,y)=\min (F_{X}(x),F_{Y}(y))$, or
equivalently, if their copula $C$ is the upper Fréchet copula $%
C(u_{1},u_{2})=\min(u_{1},u_{2}).$ Equivalently $X$ and $Y$ can be written as
nondecreasing functions of a third random variable $Z$.  As a consequence, comonotone variables maximize covariance over the set of pairs with fixed marginals:%
\begin{equation}
\mathbf{E}(XY)=\sup_{\substack{ \tilde{X}\sim X  \\ \tilde{Y}\sim Y}}\mathbf{%
E}(\tilde{X}\tilde{Y}),  \label{oneDcomonotonicity}
\end{equation}%
where $\tilde{X}\sim X$ denotes the equality in distribution between $\tilde{%
X}$ and $X$. Similarly, $X$ and $Y$ are said to have extreme negative
dependence when $X$ and $-Y$ have extreme positive dependence. Their
covariance is then minimized instead of maximized, and their copula is
the lower Fréchet copula, $C\left( u,v\right) =\max \left( u+v-1,0\right)$.

\bigskip

\noindent The present article aims at proposing an operational theory of
extreme dependence in the multivariate case, that is when $X$ and $Y$ are
random vectors. Our contribution is twofold. First, we introduce (in
Definition \ref{extreme0}) a generalization of the notion of extreme
dependence to the multivariate case, and we investigate how extreme positive
dependence generalizes in this setting. We also introduce a notion of
positive extreme dependence (in Definition \ref{extremeConic}). Then we introduce a measure of the strength of dependence based on an entropic
measure (in Section \ref{sec:IndexDep}). We then show how useful can be the concept of extreme dependence either in risk-management or in asset pricing.
 

\bigskip

\noindent \emph{Generalizing extreme dependence.} When dealing with the
multivariate case, where $X$ and $Y$ are random vectors in $\mathbf{R}^{d}$,
there is no canonical way to generalize this notion of (positive or
negative) extreme dependence and Fréchet copula. One approach, based
on the theory of Optimal Transport (see eg. the books \cite{RR:98} and \cite{Villani:2003}) would be to
consider the following optimization problem%
\begin{equation}
\max_{\substack{ \tilde{X}\sim X  \\ \tilde{Y}\sim Y}}\mathbf{E}(\tilde{X}%
\cdot \tilde{Y})  \label{optimalTransportProblem}
\end{equation}%
where $\cdot $ is the scalar product in $\mathbf{R}^{d}$. This program is a
multivariate extension of the covariance maximization problem %
\eqref{oneDcomonotonicity} and defines as extreme the distribution of the
pair $(\tilde{X},\tilde{Y})$ solution to the above problem. However it does not take into account the cross-dependence between $X_{i}$ and $Y_{j}$ for $i\neq j$. \bigskip

\noindent A more satisfactory generalization is based on the idea that both
positive and negative extreme dependence are obtained by the maximization of
a non-zero bilinear form in $\left( X,Y\right) $ over the set of couplings of
$X$ and $Y$ (i.e. joint distributions with fixed marginals). In other words,
we consider solutions of (\ref{optimalTransportProblem}), where the scalar
product is replaced by a non-zero bilinear form. This will be our notion of
\emph{multivariate extreme dependence}: random vectors $X$ and $Y$ are said to
exhibit extreme dependence if their cross-covariance matrix maximizes the
expected value of a non-zero bilinear form over all the couplings of $X$ and $%
Y$. These extreme couplings are proposed as a generalization of Fréchet
(positive and negative) extreme dependence in the multivariate case. We provide a natural geometric characterization of this notion by
considering the \emph{covariance set}, that is the set of all cross-covariance
matrices $\mathbf{E}(XY^{\prime})$ for all the couplings of $X$ and $Y$.
We show that $X$ and $Y$ have extreme dependence if and only if their
cross-covariance matrix lies on the boundary of the covariance set. \bigskip

\noindent We then turn to generalizing the notion of extreme \textit{positive%
} dependence. \ One natural way to generalize extreme positive dependence is
to look for the couplings $(\tilde{X},\tilde{Y})$ having a
cross-covariance matrix $Cov(\tilde{X},\tilde{Y})=\mathbf{E}(\tilde{X}\tilde{%
Y}^{\prime })=(\mathbf{E}(\tilde{X}_{i}\tilde{Y}_{j}))_{i,j}$ which would be
maximal for a certain partial (conical) ordering on matrices. As we
shall see, it turns out that extreme positive
dependence implies extreme dependence, and we characterize the geometric
locus of extreme positive dependent vectors on the covariance set.
\bigskip

\noindent \emph{Stress-testing dependence.} We give a method to associate
any coupling, for example any empirical coupling, with an extreme coupling,
by means of entropic relaxation technique. An algorithm is described and
results concerning its implementation are given. In particular, this algorithm provides a means to compute effectively the covariance set. We then apply these results to build stress-tests of multivariate dependence for portfolio management and to the pricing of derivatives on multiple underlyings. We also propose the construction of indices of maximal dependence, that is linear combinations of assets that have remarkable properties of extreme dependence.
\bigskip

\noindent The article is organized as follows: the first section presents
the notion of covariance set and the definition of couplings with extreme
dependence, as well a characterization of such couplings.
The second section defines and characterize couplings with positive extreme dependence, in relation to the notion of extreme dependence. The third section provides a algorithm to compute extreme couplings and the covariance set. An index of dependence, the affinity matrix is defined; a method to associate any coupling with an extreme coupling is described. We conclude with financial applications, namely stress-testing portfolio allocations and options pricing, as well as the computation of indices with extreme dependence. All proofs are collected in appendix \ref{ProofResults}.

\bigskip

\noindent \emph{Notations, definitions.}
Let $P$ and $Q$ be two probability distributions on $\mathbf{R}^{I}
$ and $\mathbf{R}^{J}$, with finite second order moments. Without
restricting the generality we assume that $P$ and $Q$ have null first
moments, so that the second order moments $\mathbf{E}(X_{i}Y_{j})$ are
indeed covariances. $\Pi (P,Q)$ is the set of all probability distributions
over $\mathbf{R}^{I}\times \mathbf{R}^{J}$ having marginals $P$ and $Q$. We
refer to an element of $\Pi (P,Q)$ as a \emph{coupling}, understating the
probabilities $P$ and $Q$. If $M$ and $N$ belong to $\mathbf{M}_{I,J}(%
\mathbf{R})$, the set of real matrices of size $I \times J$, their scalar product is denoted by $M\cdot N=Tr\left(
M^{\prime }N\right) $. If $\left( X,Y\right) \sim \pi \in \Pi (P,Q)$, we
denote indifferently $\sigma _{X,Y}$ or $\sigma _{\pi }$ the matrix with
general term $\mathbf{E}(X_{i}Y_{j})$, which is the covariance between $X_{i}
$ and $Y_{j}$; it is the cross-covariance matrix between $X$ and $Y$. Remark
that $\sigma _{X,Y}$ is the upper-right block of the variance-covariance
matrix of the vector $Z=(X,Y)^{\prime }$, and that $\ \sigma _{X,Y}$ is neither a square matrix nor a symmetric matrix in general.
\newline Moreover, we will say that a coupling $\pi$ `projects' onto $\sigma_{\pi}$, interpreting the function $\pi \mapsto \sigma_{\pi}$ as a projection operator.\bigskip

\noindent Eventually, let us recall that the \textit{subdifferential} $%
\partial f(x_0)$ of a convex function on $\mathbf{R}^n$ at a point $x_0$ is
defined as set of vectors $v$ such that $f(x) - f(x_0) \geq v \cdot (x- x_0)$
for all $x \in \mathbf{R}^n$. Here the dot is the usual scalar product. It
reduces to $\{\nabla f(x_0)\}$ if $f$ is differentiable at $x_0$, which is
true for almost every $x_0$ according to Rademacher theorem.

\section{Related literature and contribution}
As mentioned in the introduction, the extension to the multivariate setting
of the correlation maximization problem \eqref{oneDcomonotonicity} has been
tackled by several authors in order to define notions of \textit{%
multivariate comonotonicity}. \cite{puccettiScarsini}
list several possible definitions of multivariate comonotonicity, among
which two of them are directly related to the variatonal problem %
\eqref{optimalTransportProblem}. Namely, \textit{c-comonotonicity} refers to
the couplings solving problem \eqref{optimalTransportProblem}: these are the
optimal quadratic couplings of Optimal Transport Theory, also called \textit{%
maximum correlation couplings}. This variational approach to multivariate
comonotonicity is also the basis of \cite{EGH}
and \cite{GH}. They propose to extend the univariate
notion of comonotonicity and define the \textit{$\mu $-comonotonicity} by
stating that two vectors $X$ and $Y$ are $\mu $-comonotone if there exists a
random vector $U\sim \mu $ such that
\begin{equation*}
\begin{array}{ll}
\mathbf{E}(X\cdot U) & =\max \{\mathbf{E}(X\cdot \tilde{U}),\,\tilde{U}\sim
\mu \} \\
\mathbf{E}(Y\cdot U) & =\max \{\mathbf{E}(Y\cdot \tilde{U}),\,\tilde{U}\sim
\mu \}.%
\end{array}%
\end{equation*}%
This notion of comonotonicity has the advantage of being transitive, unlike
c-comonotonicity. \cite{CDG} showed that this
notion of comonotonicity appeared as `more natural' than the other ones
because it is directly related to Pareto efficiency. \bigskip

\noindent This article aims at finding multivariate couplings which exhibit
a form of strong dependence, just as the previously defined comonotonic
couplings. In what follows, the couplings defined as `extreme' are
comonotonic couplings (in the sense of the c-comonotonicity) \textit{up to a
linear transform} of one marginal (the c-comotonic coupling corresponds to
the identity transform). In other words, an extreme coupling $(X,Y)$
satisfies the variational problem
\begin{equation}
\mathbf{E}(X^{\prime }MY)=\sup_{\pi \in \Pi (P,Q)}\mathbf{E}_{\pi
}(X^{\prime }MY).  \label{variationalPbWithM}
\end{equation}%
This definition of extreme dependence is broad enough to encompass `positive
dependence' as c-comonotonicity as well as `negative dependence'
(counter-comonotonicity in the univariate case). Furthermore, it allows for
a geometrical interpretation of extreme dependence: the cross-covariance matrix of an extreme coupling is located on the boundary of the compact and convex
set of all possible cross-covariance matrices, called the covariance set. This
set has been introduced in \cite{GS} in the case with
discrete marginals, and generalized to the case with continuous marginals in \cite{DG}. Taking advantage of this simple interpretation, we then characterize the couplings $\pi $ which have a cross-covariance matrix $\sigma _{\pi }$ that are maximal for some partial orders $\succ$.\bk Although the idea of generating extreme dependence by solving problem \eqref{variationalPbWithM} arises naturally from the theory of optimal transport -- and more generally in the theory of  distributions with given marginals, see e.g. \cite{Tiit}, the computation of the covariance set remained a difficult point until now. The rest of the article proposes a method to compute extreme couplings, and for any given coupling $\hat{\pi}$, proposes a means to build a continuous sequence of couplings $\pi _{T}$ with $\pi _{0}$ being extreme, and $\sigma _{\pi_{1}}=\sigma _{\hat{\pi}}$. This is done by penalizing the problem \eqref{variationalPbWithM} with an entropy term, which allows for fast computations when the marginals are discrete distributions of probability, thanks to the Iterative Proportional Fitting Procedure. This algorithm goes back to \cite{demingStephan}, and has been used by \cite{yuille} (although they do not refer explicitly to IPFP) to solve the assignment problem, and in econometrics in \cite{GS}.

\section{Multivariate extreme dependence}

In this section we detail the notion of multivariate extreme
dependence we propose. Consider the covariance set, the set of cross-covariance matrices of couplings $\pi \in \Pi (P,Q)$:

\begin{defi}
\textit{The covariance set} $\mathcal{F}\left(P,Q\right)$ is defined as:
\begin{equation*}
\mathcal{F}\left( P,Q\right) =\left\{ \Sigma \in \mathbf{M}_{I,J}(\mathbf{R}%
):\exists \pi \in \Pi (P,Q),\Sigma _{ij}=\mathbf{E}_{\pi }(X_{i}Y_{j}),
\text{for all } i,j\right\} .
\end{equation*}
\end{defi}
\noindent As $\Pi (P,Q)$ is a convex and compact set (a proof of this property
can be found in \cite{Villani:2003}, pp. 49-50), the covariance set is also a convex compact subset of $\mathbf{M}_{I,J}(\mathbf{R})$.

\begin{figure}[tbp]
\centering
\includegraphics[width = 120 mm]{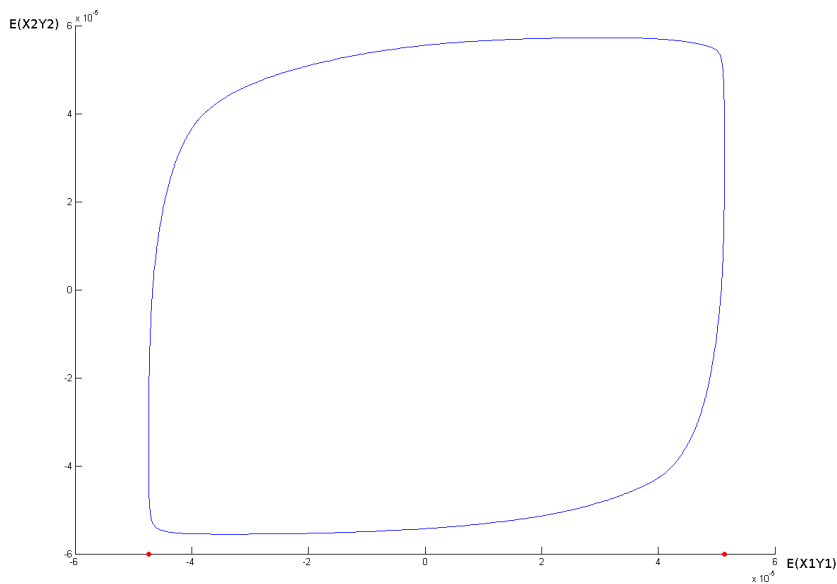}
\caption{Example of a 2 dimensional section of a covariance set}
\label{fig:simpleCovario}
\end{figure}
\bigskip \noindent Figure \ref{fig:simpleCovario} gives an example of
the 2 dimensional section of a covariance set, meaning that only the diagonal elements of the cross-covariance matrix are represented. $P$ and $Q$ are discrete distributions on $\mathbf{R}^{2}$ with equally weighted atoms
and we look at the two component-wise covariances $\mathbf{E}%
(X_{1}Y_{1})$, $\mathbf{E}(X_{2}Y_{2})$. The solid curve is the boundary of
the covariance set: every coupling between $P$ and $Q$ has a
cross-covariance matrix located within the convex hull of this curve.
The independence coupling projects on the point $(0,0)$. The dots on the $x$%
-axis represent respectively the minimal and maximal covariances between $%
X_{1}$ and $Y_{1}$. These covariances would be attained in terms of copulas by the lower and upper Fréchet copulas. This motivates our
definition of \emph{extreme dependence couplings} as couplings whose
cross-covariance matrices are on the boundary of the covariance set.

\begin{defi}
\label{extreme0}A coupling $\left( X,Y\right) \sim \pi \in \Pi (P,Q)$ has
extreme dependence if and only if $\left( \mathbf{E}_{\pi
}(X_{i}Y_{j})\right) _{ij}$ lies on the boundary of the covariance set $%
\mathcal{F}\left( P,Q\right) $.
\end{defi}
\noindent The cross-covariance matrix between $X$ and $Y$, $\sigma _{X,Y}$,
satisfies: 
\begin{equation}
Tr\left( M^{\prime }\sigma _{X,Y}\right) =\mathbf{E}(X^{\prime }MY),\ \text{%
for all}\ M\in \mathbf{M}_{I,J}(\mathbf{R})  \label{compatibility}
\end{equation}%
which allows to reformulate the notion of extreme dependence as follows:

\begin{theoreme}
\label{maxDep} The following conditions are equivalent:

i) $\left( X,Y\right) \sim \pi \in \Pi \left( P,Q\right) $ have extreme
dependence;

ii) there exists $M \in \mathbf{M}_{I,J}(\mathbf{R}) \backslash \{0\}$ such
that
\begin{equation*}
Tr\left( M^{\prime }\sigma _{\pi }\right) =\sup_{\tilde{\pi}\in \Pi
(P,Q)}Tr\left( M^{\prime }\sigma _{\tilde{\pi}}\right)
\end{equation*}%
or equivalently
\begin{equation}
\mathbf{E}_{\pi }(X^{\prime }MY)=\sup_{\tilde{\pi}\in \Pi (P,Q)}\mathbf{E}_{%
\tilde{\pi}}(X^{\prime }MY);  \label{maxCorrelPb}
\end{equation}

iii) there exists $M\in \mathbf{M}_{I,J}(\mathbf{R}) \backslash \{0\}$ and a
convex function $u$ on $\mathbf{R}^I$ such that $MY\in \partial u\left(
X\right) $ holds almost surely.
\end{theoreme}
\noindent This theorem is a corollary of the characterization of optimal couplings proved in \cite{rachevRusch:90} and \cite{Brenier:91}. 
\bk Thus a coupling $(X,Y)$ is extreme if and only if there exists a linear
transform, namely a nontrivial matrix $M$, such that $(X,MY)$ is a maximum
correlation coupling. In dimension 1, the interpretation is obvious: two
real random variables have extreme dependence iff there exists a scalar $%
m\neq 0$ and a nondecreasing function $u$ such that $mY=u(X)$. According to
the classic terminology, $X$ and $Y$ are said comonotonic if $m>0$, and
anti-comonotonic otherwise. \newline
When $M=Id$ in \eqref{maxCorrelPb}, the optimal coupling is the optimal
transport coupling for the quadratic cost (it solves problem \eqref{optimalTransportProblem}).

\section{Positive extreme dependence}

The aim of this section is to propose a generalization of the concept of Fré%
chet copula of upper dependence to the multivariate case. As already
mentioned, copula theory fails to handle this problem. Indeed, if $C_{P}$
and $C_{Q}$ are two copulas, the first in dimension $I$ (associated with
distribution $P$) and the second in dimension $J$ (associated with distribution $Q$), a natural candidate for a copula modeling positive extreme dependence would be $C_{\pi }\left( x,y\right) =\min (C_{P}(x_{1},\dots
,x_{I}),C_{Q}(x_{1},...,x_{J}))$. But according to an
`impossibility theorem' due to \cite{schweizer}, $C_{\pi}$ is a copula function
if and only if $C_{P}$ and $C_{Q}$ are themselves upper Fréchet copulas. We thus depart from the copula approach and aim at
characterizing positive extreme dependence through the
cross-covariance matrix of $X$ and $Y $. Starting from the
observation that in the univariate case, the positive extreme dependence
attains maximum covariance between $X$ and $Y$ over all the couplings of $P$
and $Q$, we introduce a conic order on the cross-covariance matrices $%
\sigma _{X,Y}$ and define positive extreme dependent couplings as the
couplings whose cross-covariance matrices are maximal with respect to this order.
\bigskip

\noindent For a given compact convex set $B \subset \mathbf{M}_{I,J}(\mathbf{R})$ such that $%
0 \notin B$ (such a set is called a compact basis), a closed convex cone in $%
\mathbf{M}_{I,J}(\mathbf{R})$ is defined by setting:
\begin{equation}  \label{basisCone}
K(B) = \{y \in \mathbf{M}_{I,J}(\mathbf{R}) \vert x \cdot y \geq 0,\,
\forall x \in B\}
\end{equation}
Considering cones of this form might seem restrictive (appendix \ref{FactsConic} gives more details on such cones), yet we provide below some
examples that show that classic cones can be defined in such a manner.
\bigskip

\noindent
Let $M_1$, $M_2$ be two matrices in $\mathbf{M}_{I,J}(\mathbf{R})$. A \textit{strict} conic order on $\mathbf{M}_{I,J}(\mathbf{R})$ is defined by
\begin{equation*}
M_1 \succ_{K(B)} M_2 \ \text{if} \ M_1 - M_2 \in Int(K(B))
\end{equation*}
The interior of $K(B)$ is $\{y \in \mathbf{M}_{I,J}(\mathbf{R}) \vert x
\cdot y > 0,\, \forall x \in B\}$.

\begin{defi}
\label{extremeConic} Let $B$ be a compact basis. A coupling $(X,Y)$ such that $\sigma _{X,Y}$ is a maximal element in $\mathcal{F}\left( P,Q\right) $ with respect to the
strict conic order $\succ_{K(B)}$\ is said to have \emph{positive extreme
dependence} with respect to $\succ _{K(B)}$.
\end{defi}

The following results fully characterize couplings with positive extreme
dependence in terms of maximal correlation couplings.

\begin{theoreme}
\label{maxUpperDep}The following conditions are equivalent:

i) $\left( X,Y\right) \sim \pi \in \Pi \left( P,Q\right) $ have extreme
positive dependence with respect to $\succ_K(B)$

ii) there exists $M\in B$ such that
\begin{equation*}
Tr\left( M^{\prime }\sigma_{\pi }\right) =\sup_{\tilde{\pi} \in \Pi
(P,Q)}Tr\left( M^{\prime }\sigma_{\tilde{\pi}}\right)
\end{equation*}%
or equivalently
\begin{equation}  \label{maximalityCharac}
\mathbf{E}_{\pi }(X^{\prime }MY)=\sup_{ \tilde{\pi}\in \Pi (P,Q)}\mathbf{E}_{%
\tilde{\pi}}(X^{\prime }MY)
\end{equation}

iii) there exists $M \in B$ and a convex function $u$ such that $MY\in
\partial u\left( X\right) $ holds almost surely
\end{theoreme}
\noindent Hence, $\sigma _{X,Y}$ is maximal if and only if there exists $M\in C$ such that $X$ and $MY$ are maximally correlated. Obviously, this result is a close parallel to Theorem \ref{maxDep} except that $M$ is constrained to belong to $B$. As a consequence the positive extreme couplings are a particular case of extreme couplings. The interpretation in dimension 1 is again straightforward: $X$ and $Y$ have positive
extreme dependence (w.r.t. the usual order on $\mathbf{R}$) iff they are
comonotonic.

\begin{figure}[ht]
\centering
\includegraphics[width = 120 mm]{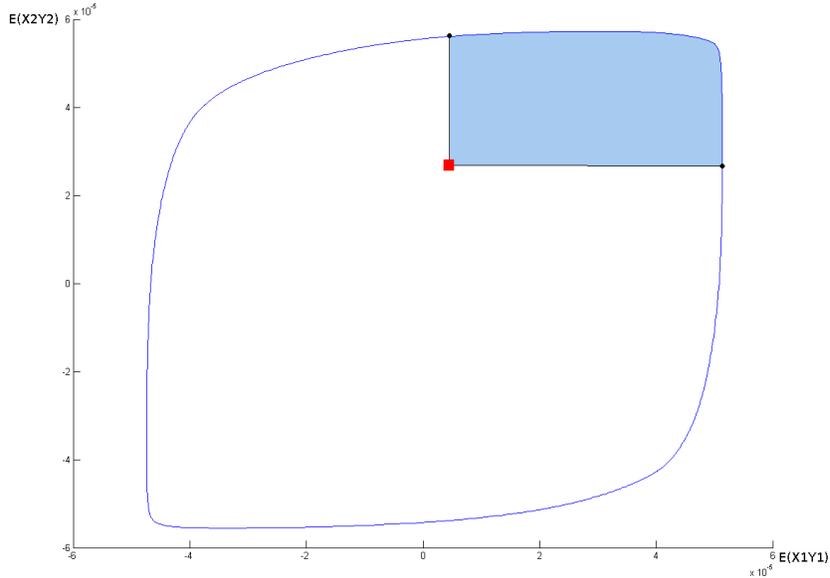}
\caption{Shaded region: location of the couplings dominating the coupling
materialized by the square dot.}
\label{fig:covario1stSight}
\end{figure}

\begin{figure}[ht]
\centering
\includegraphics[width = 120 mm]{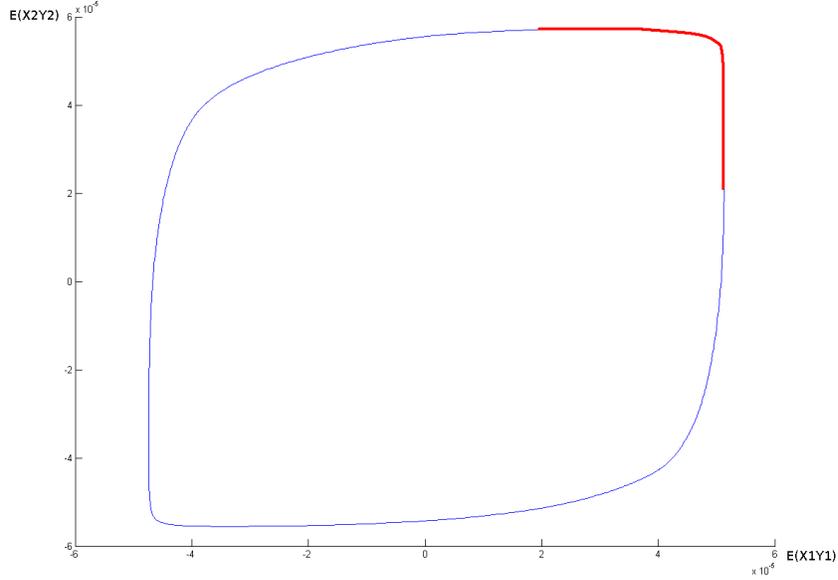}
\caption{Maximal couplings on the boundary.}
\label{fig:covarioHull}
\end{figure}
\bigskip

\noindent To better understand the relation between Definition \ref{extreme0}%
\ and Definition \ref{extremeConic}, let us go back to the two dimensional\
section of the covariance set discussed in the previous section, and consider that $K(B)$ is the positive orthant of $\mathbf{R}^{2}\times \mathbf{R}^{2}$. The shaded region in Figure \ref{fig:covario1stSight} is the set of couplings
dominating the coupling that projects on the square dot, with respect to
the orthant order; as a consequence this coupling can not have positive extreme
dependence. This intuitively explains why maximal elements should be on the
boundary of the covariance set, hence that positive extreme couplings should be
extreme couplings. Maximal elements are represented on the bold curve in figure  \ref{fig:covarioHull}. Consequently the couplings exhibiting positive extreme dependence project on this bold portion of the boundary of the covariance set. They form only a small part
of the couplings of extreme dependence. \bigskip

\noindent To demonstrate the applicability of this approach, here are
three examples of partial orders on covariance matrices.

\begin{ex} \textit{Orthant order}.
Let $M_{I,J}^{+}(\mathbf{R})$ (resp. $M_{I,J}^{++}(\mathbf{R})$) denotes the
set of real $I\times J$ matrices with nonnegative coefficients (resp.
positive coefficients). The set $B=M_{I,J}^{+}(\mathbf{R})\cap
\{\sum_{i,j}M_{i,j} = 1\}$ is a compact basis. $K(B)$ is easily seen to
be the set $M^+_{I,J}(\mathbf{R})$ and its interior is $M_{I,J}^{++}(\mathbf{%
R})$.  Eventually $M_1\succ M_2$ iff $M_1-M_2$ has only positive coefficients: this
is the (strict) \emph{orthant order} on matrices.
\end{ex}

\begin{ex}\textit{Loewner order}.
\label{ex:Loewner}Let $S_{n}^{+}$ and $S_{n}^{++}$ denote respectively the
set of nonnegative matrices in $S_{n}$ and the set of definite positive
matrices in $S_{n}$. If $B=\{S\in S_{n}^{+}(\mathbf{R})|Tr(S)=1\}$ is the
set of semi-definite matrices with unit trace, $B$ is a convex compact
subset of $\mathbf{M}_{n}(\mathbf{R})$ and $K(B)=\{M\in \mathbf{M}_{n}(%
\mathbf{R})|Tr(M^{\prime }S)\geq 0,\forall S\in B\}$ is the set of matrices $%
M$ whose symmetric part, $\frac{M+M^{\prime }}{2}$, is semi-definite
positive. The strict order $\succ _{K(B)}$ is then defined as: $M_1\succ M_2$
iff the symmetric part of $M_1-M_2$ is definite positive. This is an extension
to $\mathbf{M}_{n}(\mathbf{R})$ of the classic \emph{Loewner order} on
symmetric matrices.
\end{ex}
\noindent The following trivial example shows that the ordering induced by Example \ref{ex:Loewner}\ allows various positive extreme couplings. A first remark is that the maximum correlation coupling is indeed positive extreme, by setting $M=Id$ in theorem \ref{maxUpperDep}. Consider $P\sim \mathcal{N}(0,I_{2})$, the bivariate normal distribution, and $Q=\mathcal{N}(0,1)\otimes \mathcal{U}_{(0,1)}$, the distribution of a vector whose first component is normal and the second one is the uniform distribution on $(0,1)$, independent from the first component. Let $X\sim P$ and $Y=(X_{1},U)^{\prime }$, $U\sim \mathcal{U}_{(0,1)}$ independent from $(X_{1},X_{2})$, so that $Y\sim Q$. This coupling has not the maximum correlation even though $X_{1}=Y_{1}$. However it satisfies (\ref{maximalityCharac}) with $A=\left(
\begin{smallmatrix}
1 & 0 \\
0 & 0%
\end{smallmatrix}%
\right) $ and qualifies as a maximal coupling.

\section{An index of dependence\label{sec:IndexDep}}

Suppose now we are observing (or simulating) a coupling $\hat{\pi}\in \Pi (P,Q)
$, thereafter referred to as an empirical coupling. Even if this coupling is supposed to exhibit strong dependence, its cross-covariance matrix will never be exactly
located on the boundary of the covariance set. Our problem is then to \emph{%
associate an extreme coupling with $\hat{\pi}$}; more precisely, we propose
to find a continuous sequence of non deterministic couplings $\pi_{T}$ such
that $\pi _{1}=\hat{\pi}$ and $\pi _{0}$ is an extreme coupling. In other
words, we give a means to go smoothly from an empirical coupling to an
extreme one by progressively increasing the strength of the dependence
between the marginals.
 This is done by introducing an entropic penalization
of \eqref{maxCorrelPb}, so that its solutions project on inner points of
the covariance set.

\subsection{Entropic relaxation}
\label{entropicRelaxation}
Consider the following problem, which is the entropic penalization of %
\eqref{maxCorrelPb}:
\begin{equation}  \label{EntropyProblem}
W(M,T):= \max_{\pi \in \Pi(P,Q)} \left(\mathbf{E}_{\pi}(X^{\prime}MY) + T
Ent(\pi)\right)
\end{equation}
The entropy of a coupling $\pi$ is defined as
\begin{equation*}
Ent(\pi) = \left\{%
\begin{array}{l}
-\int \log \pi(x,y) d\pi(x,y),\, \text{{\footnotesize if $\pi \ll dx \otimes
dy$ and the integral exists and is finite}} \\
- \infty \ \text{otherwise}%
\end{array}
\right.
\end{equation*}
The parameter $T$ can be thought of as a `temperature' parameter which controls the strength of the entropic penalization. The problem \eqref{maxCorrelPb} corresponds to $T = 0$, while letting $T$ to $+\infty$ amounts to maximize the entropy, in which case the solution of problem \eqref{EntropyProblem} is the independence coupling.\newline 
Let $\pi_{M,T}$ denote a solution of \eqref{EntropyProblem}; a proof of its
existence can be found in \cite{Ruschendorf:95} and references
therein. We assume furthermore that the entropy of $\hat{\pi}$ is finite.

\noindent Fixing the temperature at 1, our aim in the first place is to find a
matrix $M$ such that $\hat{\pi}$ and $\pi_{M,1}$ have the same cross-covariance matrix: $%
\sigma_{\hat{\pi}} = \sigma_{\pi_{M,1}}$. By a property of the subdifferential of a maximum function, the gradient of $W$ with respect to $M$ is: $\nabla_M W(\cdot,1) = \sigma_{\pi_{M,1}}$. This
implies that $M$ is the solution of the following variational problem
\begin{equation}  \label{minProblem}
\min_{M \in M_{I,J}(\mathbf{R})} W(M,1) - \sigma_{\hat{\pi}} \cdot M
\end{equation}
$W(\cdot,1)$ is a convex function as a supremum of affine functions in $M$,
and consequently the objective function in \eqref{minProblem} is convex as
well: this is a classic unconstrained convex minimization problem. Moreover, \eqref{minProblem} is bounded below, which yields the existence of a global minimizer. A detailed proof and a discussion of uniqueness in \eqref{minProblem} is given in appendix \ref{detailsMinProblem}.\bk
\begin{figure}[htbp]
\centering
\includegraphics[width = 120 mm]{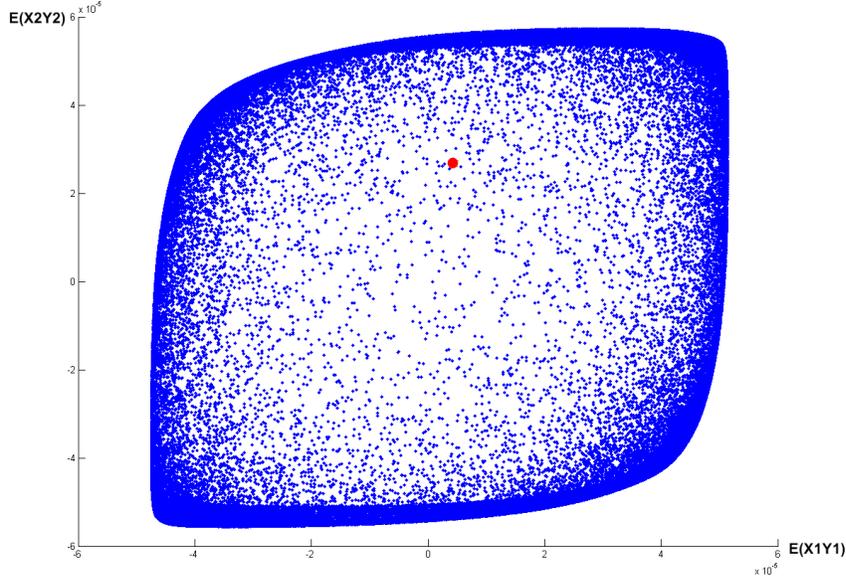}
\caption{Projection of various $\protect\pi_M$}
\label{fig:basic_covario}
\end{figure}
\noindent Figure \ref{fig:basic_covario} shows the diagonal of $\sigma_{\pi_{M,1}}$ in the coordinates $(\esp(X_1Y_1),\esp(X_2Y_2))$, for a large number of randomly generated matrices $M$. This graph is obtained by sampling many matrices $M$ with coefficients uniformly distributed in the interval $[-1,1]$, and then solving for each simulated M the problem \eqref{EntropyProblem}, in order to find $\sigma_{\pi(M,T)}$. $T$ is taken small enough to obtain near from extreme couplings. The solution of  \eqref{EntropyProblem} is computed thanks to the algorithm presented in section \ref{numericalSolution}. The bullet point has coordinates $(\esp_{\hat{\pi}}(X_1Y_1),\esp_{\hat{\pi}}(X_2Y_2))$. One sees that any inner point of the covariance set can be attained by a properly chosen $\pi_M$. This is a noticeable advantage of the entropic relaxation: not only the optimal couplings solving \eqref{EntropyProblem} are easily computed (at least when the marginals are discrete, see section \ref{numericalSolution}), but changing the temperature parameter allows to reach any cross-covariance matrix inside the covariance set.

\subsection{Numerical solution}
\label{numericalSolution}
The optimal $\pi_{M,1}$ in \eqref{EntropyProblem} obeys
the following equation (see \textit{e.g.} \cite{Ruschendorf:95} for a proof):
\begin{equation*}
\begin{array}{ccc}
\log \pi_{M,1}(x,y) & = x^{\prime}My + u(x) + v(y), & u \in L^1(dP), v \in
L^1(dQ)%
\end{array}%
\end{equation*}
In other words, the log-likelihood of $\pi_{M,1}$ is the sum of a quadratic
term $x^{\prime}My$ and of an additively separable function in $x$ and $y$. The
solution is found by setting $u$ and $v$ such that $\pi_{M,1}$ has the
marginals $P$ and $Q$. This is the purpose of the Iterative Projection Fitting Algorithm (\cite{demingStephan}, Von Neumann 1950).
\bk
This algorithm consists in building a sequence $\pi_n$ such that $\pi_{2n}$ has first marginal $P$ and $\pi_{2n +1}$ has second marginal $Q$. It turns out that $\pi_n$ converges towards a probability $\pi$ with correct marginals $P$ and $Q$. \newline
When the marginals $P$ and $Q$ are discrete distributions with atoms $P(x)$ and $Q(y)$ respectively, the algorithm is straightforward, as it consists in solving a series of linear systems:
\begin{equation*}
\left\{
\begin{array}{ccc}
e^{v_{n+1}(y)} & = & \frac{Q(y)}{\sum_{x} r(x,y)e^{u_n(x)}} \\
e^{u_{n+1}(y)} & = & \frac{P(x)}{\sum_{y} r(x,y)e^{v_{n+1}(y)}}%
\end{array}
\right.
\end{equation*}
where $r(x,y) = \frac{e^{x^{\prime}My}}{\sum_{x,y}e^{x^{\prime}My}}$.
\newline The convex unconstrained minimization problem \eqref{minProblem} can be solved by a Quasi-Newton algorithm (we used the BFGS method in the examples below).
\newline Of course, this algorithm can be used for any temperature $T$, by replacing $M$ by $\frac{M}{T}$ in the previous equations.

\subsection{Derivation of the extreme coupling}

\label{derivationExtremeCoupling} We recall that our aim is to associate an
inner coupling (i.e. a coupling whose cross-covariance matrix is inside the covariance set) to some extreme coupling which projects onto the boundary of the covariance set, by finding a trajectory of couplings that goes smoothly from the inner one to the extremal one. \bigskip

\noindent The previous algorithm yields a particular matrix $\hat{M}$ and a
coupling $\pi_{\hat{M}}$ such that $\sigma_{\hat{\pi}} = \sigma_{\pi_{\hat{M}%
,1}}$. This coupling was found by setting arbitrarily the temperature at 1;
the entropy penalization was thus effective and this allowed to reach inner
points in the covariance set. This temperature parameter is easily explained.
When it goes to $+\infty$, the entropy penalization is predominant in %
\eqref{EntropyProblem}. Intuitively, the solution is the coupling
with maximal `disorder': this is the independence coupling. On the
contrary, the less is the temperature, the closer \eqref{EntropyProblem} is
to the non penalized problem. Hence, the lower $T$, the more $\pi_{\hat{M},T}
$ projects near the boundary of the covariance set. Hence associating $\hat{\pi}$ with an extreme
coupling can be done in the following way: once $\hat{M}$ is found, a
sequence of $\pi_{\hat{M},T_n}$, $T_n \downarrow 0$ yields a trajectory of cross-covariance matrices which tends to the boundary.\bk
\begin{figure}[htbp]
\centering
\includegraphics[width = 120 mm]{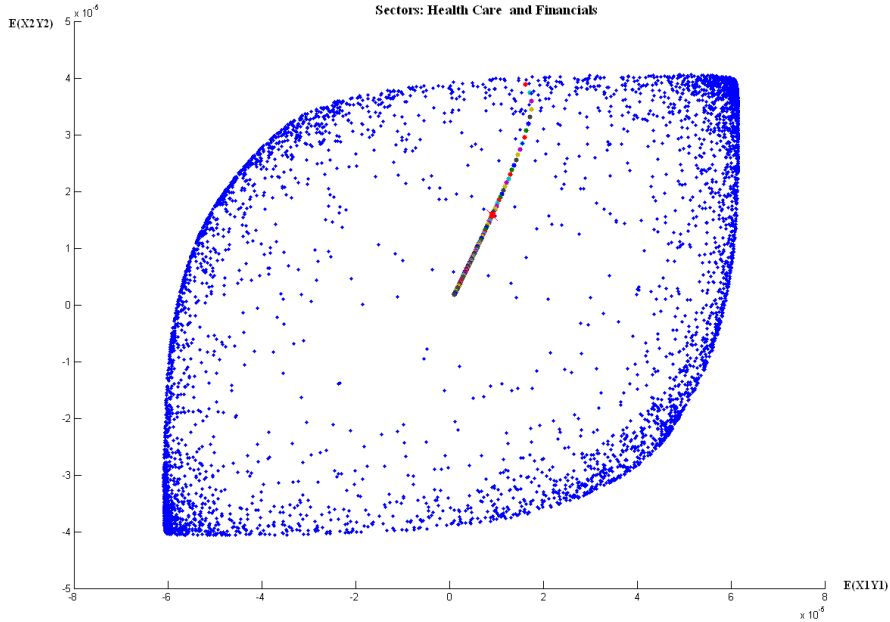}
\caption{A trajectory toward an extreme coupling when the sectors are Health
Care and Financials}
\label{fig:trajectory1}
\end{figure}
Figure \ref{fig:trajectory1} summarizes this idea: each point on the curve
is the projection of some $\pi_{\hat{M},T_n}$. As $T \to +\infty$, we
recover the independence coupling whose projection is located at (0,0). When
the temperature decreases, the trajectory passes on $\hat{\pi}$ at $T = 1$,
and gradually approaches the boundary of the covariance set. The entropy is
decreasing along this trajectory, as $Ent(\pi_T)$ decreases as $T \downarrow
0$ (thanks to the convexity of $W(M,T)$ in $T$), and thus lowering the
temperature corresponds to going away from the independence coupling
(maximal entropy). Thus, the temperature can be seen as a means to control
the strength of the dependence. The matrix $\hat{M}$ can be seen as an
affinity matrix: in the limit of $T \to 0$, the extreme coupling $\pi_{\hat{M%
},0}$ achieves the supremum of $\mathbf{E}_{\pi}(X^{\prime}\hat{M}Y)$. Thus $%
\hat{M}$ is the linear transform that makes $X$ the most dependent with $%
\hat{M}Y$ under $\pi_{\hat{M},0}$. \bigskip

\noindent This can be used to define formally an \emph{index of dependence},
for $\hat{\pi}$ different from the independence coupling: choosing a norm $%
||\cdot ||$ over the set of matrices $M_{I,J}(\mathbf{R})$ and using the
homogeneity of $W$, namely $W(\lambda M,\lambda T)=\lambda W(M,T)$ for all $%
\lambda \in \mathbf{R}$, we have $\pi _{\hat{M},1}=\pi _{\hat{M}/||\hat{M}%
||,1/||\hat{M}||}$ and the temperature $\frac{1}{||\hat{M}||}$ appears as an
indicator of the strength of the dependence between the marginals of $\hat{%
\pi}$.
%
%
%
%

\section{Applications}
\label{introApplication}
In the following financial applications below, we use the technique described in the previous section with times series of linear daily returns on industrial sectors of mainstream indices: S\&P 500 and DJ Eurostoxx.
We consider Health Care, Financial and Food \& Beverage sectors of these
indices: $P$ and $Q$ are distributions on $\mathbf{R}^{3}$. The historical
data spans 5 years between September 2004 and September 2009. Table \ref%
{summaryStatistics} gives summary statistics
\begin{table}[tph]
\caption{Summary Statistics}
\label{summaryStatistics}\centering
\begin{tabular}{ll}
\toprule Mean Returns & $10^{-4 }\left(%
\begin{smallmatrix}
1.03 & -1.13 & 1.67 & 1.16 & -1.37 & 3.99%
\end{smallmatrix}
\right)$ \\
Variance & $10^{-4}.\left(%
\begin{smallmatrix}
1.36 & 7.65 & 1.16 & 1.14 & 4.15 & 1.12%
\end{smallmatrix}
\right)$ \\
Correlation matrix & $\left(%
\begin{smallmatrix}
1 &  &  &  &  &  \\
0.66 & 1 &  &  &  &  \\
0.76 & 0.62 & 1 &  &  &  \\
0.22 & 0.10 & 0.19 & 1 &  &  \\
0.26 & 0.33 & 0.25 & 0.49 & 1 &  \\
0.22 & 0.16 & 0.22 & 0.67 & 0.58 & 1.00%
\end{smallmatrix}%
\right) $ \\
&  \\
Cross-Covariance & $10^{-5}.\left(%
\begin{smallmatrix}
2.74 & 3.05 & 2.13 \\
6.04 & 1.8 & 5.52 \\
2.66 & 4.62 & 2.56%
\end{smallmatrix}%
\right)$ \\
\bottomrule &
\end{tabular}%
\end{table}
(the three first variables corresponds to S\&P sectors, the last third to
Eurostoxx). In particular, the correlations between sectors belonging to
different indices are mild ($<35\%$). Inside an index,
correlation is well higher, but remains below 80\%; this motivated our
choice for these sectors: the marginal distributions are not degenerated.

\subsection{Numerical Results}

$P$ and $Q$ are discrete distributions with equally weighted atoms in $%
\mathbf{R}^3$, each atoms being a vector of the returns at some date of the
three sectors.
\begin{equation*}
P = \frac{1}{N} \sum_{t = 1}^N \delta_{r^X_t}\quad , \quad r^X_t = \text{%
{\small {vector of the linear returns at date $t$ on the three sectors of the S\&P500}}}
\end{equation*}
The optimal $\hat{M}$ we find when considering the three sectors or the
Construction and Health Care sectors are:

\begin{center}
\begin{tabular}{lcc}
\toprule \# of components & 2 & 3 \\
\midrule optimal M & $\left(%
\begin{smallmatrix}
0.23 & -0.14 \\
-0.10 & 0.40%
\end{smallmatrix}%
\right)$ & $\left(%
\begin{smallmatrix}
0.25 & -0.139 & -0.37 \\
-0.39 & 0.44 & -0.80 \\
-0.57 & -0.15 & 0.86 \\
&  &
\end{smallmatrix}%
\right)$ \\
error = $\frac{\vert \vert \sigma_M - \sigma_{\hat{\pi}} \vert \vert}{\vert
\vert \sigma_{\hat{\pi}} \vert \vert}$ & $\approx 0.1\%$ & $< 0.2\%$ \\
\bottomrule &  &
\end{tabular}
\end{center}

The linear returns are expressed in percentage. The error is computed as the
percentage of difference between $\sigma_{\hat{\pi}}$, the cross-covariance
targeted, and $\sigma_{\pi_{M,1}}$, the covariance matrix of the optimal
coupling. They should be perfectly equal in theory and this percentage
measures the convergence of the gradient algorithm.

\subsection{Financial applications}

\noindent First, we use the trajectory of couplings
$T \mapsto \pi_{\hat{M},T}$ as a continuous family of scenarios of
increasing dependence. Theses scenarios are used to build scenarios of stress-tests involving multivariate variables, with obvious applications to risk management. By stress-testing, we mean increase the index of dependence defined above (that is, lowering the temperature parameter), thus shifting away continuously from some coupling $\hat{\pi}$ to the extreme coupling $\pi_{\hat{M},0}$. This is to be compared to the method that consists in selecting the maximum correlation coupling as the `strongest dependence scenario'; indeed this coupling might be less in line with the cross-covariance structure of the empirical coupling $\hat{\pi}$, yielding unexpected and undesired results when managing risky portfolios or options on several assets. \bigskip

\indent Then, we exploit further the affinity matrix $\hat{M}$ in order to 
exhibit indices of maximal correlation, based on an analysis of its singular value decomposition.

\subsubsection{Portfolios stress-testing}

In order to underline the necessity of accounting properly for the
multivariate dependence, the problem of one-period portfolio allocation is considered. Suppose an investor chooses to allocate his wealth between assets $X_1,\ldots, X_n,Y_1,\ldots,Y_m$. The problem is to study the impact of the change of the dependence between $X = (X_1,\dots,X_n)$ and $Y =(Y_1,\dots,Y_m)$ on the investor's portfolio. 
\newline In the numerical examples below, the assets are S\&P Sector Indices: $X$ is composed of Materials, Construction and Retail indices, while $Y$ is composed of Food and Beverage, Health Care, Financials and Utilities indices. The corresponding summary statistics are given in table \ref{summaryStatisticsII}. Correlation is higher than in the above examples as the sectors are industrial sectors on a single index, the S\&P500.

\begin{table}[htp]
\caption{Summary Statistics}
\label{summaryStatisticsII}\centering
\begin{tabular}{ll}
\toprule Mean Returns & $10^{-4}.\left(%
\begin{smallmatrix}
2.89 & 1.67 & 1.03 & -1.13 & 1.97 & 2.01 & 1.85%
\end{smallmatrix}
\right)$ \\
Variance & $10^{-4}.\left(%
\begin{smallmatrix}
3.59 & 1.16 & 1.36 & 7.65 & 1.92 & 0.984 & 3.25%
\end{smallmatrix}
\right)$ \\
Correlation matrix & $\left(%
\begin{smallmatrix}
1 &  &  &  &  &  &  \\
0.72 & 1 &  &  &  &  &  \\
0.71 & 0.76 & 1 &  &  &  &  \\
0.69 & 0.86 & 0.65 & 1 &  &  &  \\
0.69 & 0.85 & 0.69 & 0.76 & 1 &  &  \\
0.69 & 0.67 & 0.75 & 0.62 & 0.66 & 1 &  \\
0.70 & 0.76 & 0.60 & 0.72 & 0.74 & 0.56 & 1 \\
&  &  &  &  &  &
\end{smallmatrix}%
\right) $ \\
Cross-Covariance & $10^{-4}.\left(%
\begin{smallmatrix}
1.41 & 1.53 & 3.62 & 1.85 \\
0.921 & 0.979 & 1.83 & 1.05 \\
1.27 & 1.45 & 3.73 & 1.50%
\end{smallmatrix}%
\right)$ \\
\bottomrule &
\end{tabular}%
\end{table}

\bigskip

\noindent It is assumed the investor chooses his portfolio allocation according to the  Markowitz allocation problem (over
a one year horizon), meaning that the weights $\omega$ that determine the allocation are chosen by solving the problem $\max_{\sum_i \omega_i = 1}\mu \cdot \omega - \frac{\lambda}{2}\omega^{\prime}\Sigma \omega$. $\mu$ are the expected yearly returns of the stocks, $\Sigma$ the covariance matrix of the returns and $\lambda$ a risk aversion parameter specific to the investor. We assume that
both $\mu$ and $\Sigma$ are the standard empirical estimators computed over a period of one-year, the in-sample period. The risk aversion parameter $\lambda$ is set at 3. The solution to the Markowitz allocation problem with these parameters is denoted $w$. The risk of a portfolio is here identified to its variance, and is known as soon as the covariance between the assets is specified. When performing the allocation at time 0, the investor is expecting a risk of $\omega^{\prime}\Sigma w $. The dependence stress-test consists in considering that the market conditions changes after the investment decision: the strength of dependence between $X$ and $Y$ increases. \bigskip

\begin{figure}[htp!]
\centering
\includegraphics[width = 127 mm]{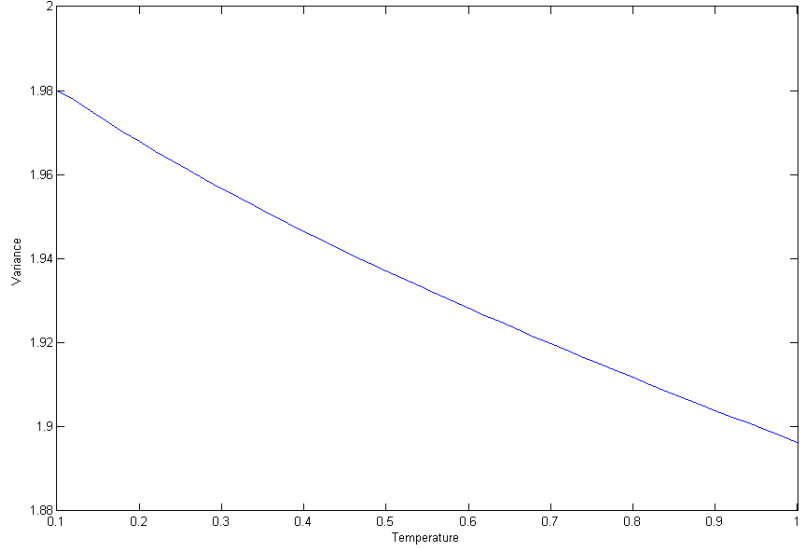}
\caption{Plot of $T \mapsto w^{\prime}\Sigma_T w$}
\label{stressed_alloc}
\end{figure}

\noindent The affinity matrix is computed with respect to the in-sample
data. The whole trajectory of couplings toward the boundary obtains,
parameterized by the temperature $T$. These couplings $\pi_T$ yield stressed
covariance matrices $\Sigma_T = \mathbf{E}_{\pi_T}((X - \mathbf{E}(X))(Y -
\mathbf{E}(Y))^{\prime})$. $\Sigma_T$ represents a scenario where the
marginals of $X$ and $Y$ are left unchanged, while the realized dependence
between $X$ and $Y$ has increased, compared to the initial covariance matrix
$\Sigma$. 
\newline The unexpected risks the investor might face when the dependence varies is materialized by the variance $w^{\prime}\Sigma_T w$, plot on graph \ref{stressed_alloc}.
The variance obtained at temperature 1 is $w^{\prime}\Sigma w$; in the worst
case (which corresponds to temperature $0.1$ on graph\ref{stressed_alloc}), the investor chooses
a portfolio that has a variance 4\% higher than expected. \newline When the
dependence is properly accounted for, the investor determines the optimal
weights $w_T$ according to the covariance $\Sigma_T$. The opportunity cost $%
\mu \cdot w_T - \mu \cdot w$ is the loss on the return  when
the dependence increases while the investor sticks to the initial allocation $w$.
This cost is more and more significant as the temperature lowers, reaching
6\% in this case.
\begin{figure}[ht]
\centering
\includegraphics[width = 120 mm]{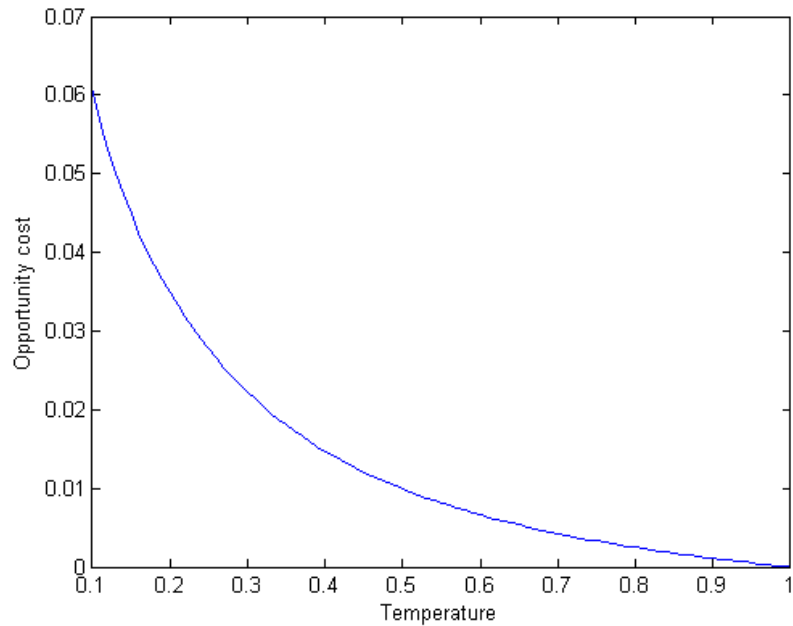}
\caption{Opportunity cost as a function of the temperature}
\end{figure}
A comparison with  the maximum correlation coupling is enlightening. First of all, this coupling is
not defined when the dimension of $X$ and $Y$ are different. Consequently an
asset is removed from $Y$ (namely the Food and Beverages index) and the same computations as above are performed: a
covariance matrix $\Sigma_B$ that would be the realized covariance if the
assets were in maximum correlation dependence is computed. On this
particular example, the variance $w^{\prime}\Sigma_B w$ is 60\% \textit{lower%
} than the expected variance $w^{\prime}\Sigma w$. Other examples can yield a significantly higher covariance. This shows that the maximum
correlation coupling might not be always adapted as a means of
stress-testing the dependence.

\bigskip \noindent A more classical way to stress the dependence
is to suppose that the correlation between $X_i$ and $Y_j$ is fixed and equal to some parameter $\rho$ for all
$i$ and $j$; the resulting cross-covariance matrix is denoted $\Sigma_{\rho}$. A problem of this method is that it is known beforehand that, depending
on the marginals, $\Sigma_{\rho}$ might not be an admissible
cross-covariance matrix for $P$ and $Q$; the resulting variance-covariance
matrix of the vector $(X,Y)$ might fail to be semi-definite positive. This
stress-test yields in this case underestimated risks. Indeed, while in our
framework the variance $w^{\prime}\Sigma w$ is at 1.91, this level of
variance is attained only when $\rho$ is above 95\%, while the mean of the
empirical cross-correlation is around 60\%. Furthermore, even if $\rho$ is
set at $100\%$ (disregarding the admissibility problem evoked above), the
resulting variance is still lower than the one obtained with the extreme
coupling. \bigskip

\noindent It appears that the trajectory $T \mapsto \pi_T$ provides a
coherent sequence of covariance matrices $\Sigma_T$ that models an increase
of the dependence between $X$ and $Y$. This method respects both marginals
and has the advantage of generating admissible matrices contrary to the usual
method of parameterizing correlation matrices by a single parameter. Moreover, the maximum correlation coupling fails in this setting to properly account for increasing the risk of dependence, likely because it ignores the cross-correlation effects.

\subsubsection{Options pricing}
These couplings with increasing strength of dependence can be also used
for the risk management and pricing of rainbow options (options on several underlyings). As a case study, consider the underlyings $X_1, \dots, X_n$, $Y_1,\dots, Y_m$. It is assumed that each one follows a log-normal martingale diffusion (i.e. we assume a null risk free rate and write the risk-neutral dynamics):
\begin{equation*}
\left\{
\begin{array}{ccc}
\frac{dX^i_t}{X^i_t} & = \sigma^X_i dW^i_t \quad,\quad d\langle
W^i,W^j \rangle_t = \rho^X_{ij}dt &\quad,\quad X^i_0 = 1\\
\frac{dY^i_t}{Y^i_t} & = \sigma^Y_i dB^i_t \quad,\quad d\langle
B^i,B^j \rangle_t = \rho^Y_{ij}dt&\quad,\quad Y^i_0 = 1
\end{array}
\right.
\end{equation*}
The model is fully specified as soon as the correlation matrix between $W$
and $B$ is set. Consider the option that pays $\Phi = \min((\max_i X^i_T -
K)_+,(\max_j Y^j_T - K)_+)$; it is the minimum between the payoffs of two
best-of options on the $X^i$ on the one hand and the $Y^j$ on the other
hand. It pays when the $X^i_T$ and $Y^i_T$ perform well, but mitigates the
gain by selecting the lowest payoff between $(\max_i X^i_T - K)_+$ and $%
(\max_j Y^j_T - K)_+$. \bigskip

\noindent Suppose an investor has sold this option and knows the distribution of the vector $X$ and $Y$. In other words, he has been able to calibrate the volatilities $\sigma^X_i$ and $\sigma^Y_i$, as well as the correlation matrices of $(W^1,\ldots,W^n)$ and of $(B^1,\ldots,B^m)$. The investor may have a guess on the dependence between $X$ and $Y$ (or equivalently between $B$ and $W$), for instance an empirical estimation of the covariance matrix, but this guess is not sufficient to price the claim $\Phi$ in a conservative manner. A way to do this is to compute the price of this claim when the strength of the dependence between $X$ and $Y$ varies from the independence coupling to some extreme coupling and pick the highest value for the claim.\bigskip

\noindent For the purpose of numerical computations, the terminal distribution of the underlyings is discretized. The atoms of the discretized marginals are respectively denoted $x^i_T$ and $y^j_T$. For each specification
of a cross-covariance matrix $A$ between $X$ and $Y$, a trajectory $\pi_T(A)$
is obtained. The claim is priced as the expected value of $\Phi$ under the distribution $\pi_T(A)$:
\begin{equation*}
\begin{split}
P_T(A) &= \mathbf{E}_{\pi_T(A)}\Big(\min((\max_i X^i_T - K)_+,(\max_j Y^j_T
- K)_+)\Big) \\
&= \sum_{x^i_T,y^j_T} \min((\max_i x^i_T - K)_+,(\max_j y^j_T -
K)_+)\pi_T(A)(x^i_T,y^j_T)
\end{split}%
\end{equation*}
In the following example, $X$ has 3 components and $Y$ has 4, $\sigma^X = (0.15,0.20,0.22)^{\prime}$ and $\sigma^Y = (0.13,0.10,0.16,0.18)^{\prime}$. For the sake of the exposition $W$ and $B$ are standard Brownian
motions ($\rho^X = Id_n$ and $\rho^Y = Id_m$) while the cross-correlation
matrix between $W$ and $B$ is randomly generated, and set at
\begin{equation*}
\left(%
\begin{smallmatrix}
0.087 & 0.126 & 0.068 & 0.100 \\
0.490 & 0.438 & 0.006 & 0.149 \\
0.136 & 0.369 & 0.447 & 0.331 \\
&  &  &
\end{smallmatrix}%
\right)
\end{equation*}
The strike is set at 1, i.e. at time 0 the option is at-the-money.

\begin{figure}[h!]
\centering
\includegraphics[scale = 0.90]{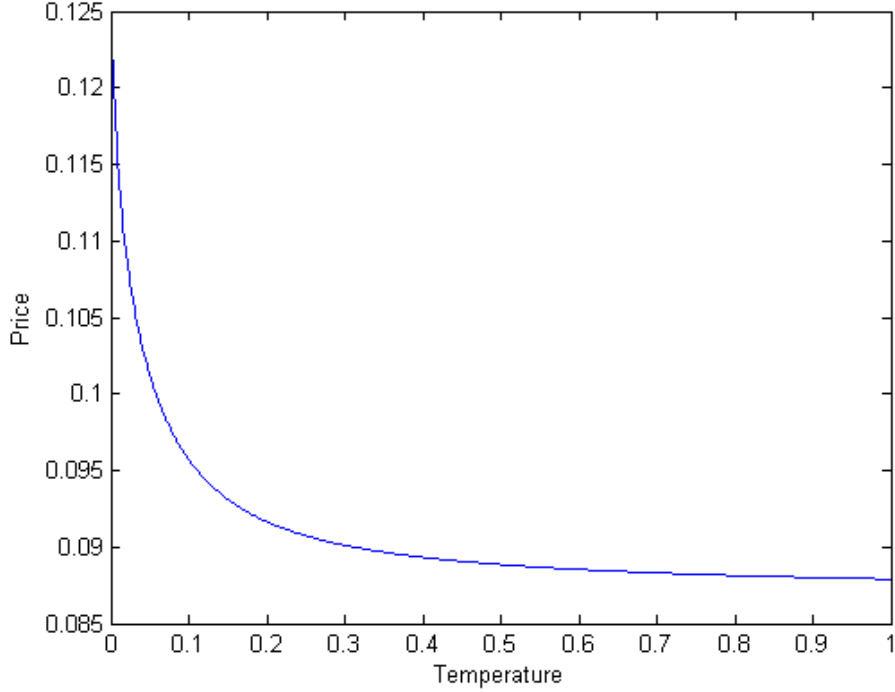}
\caption{Price as a function of the temperature}
\label{priceBasketOption}
\end{figure}
\bigskip

\noindent As seen on graph \ref{priceBasketOption}, the price increases as
the temperature lowers; this is an expected behavior, as when the dependence
between the assets increases, so does the dependence between their
respective maxima and hence the minimum of these maxima tends to be higher,
which yields a higher price. In this setting, the stress-test increases the
price by more than 30\% (i.e. between the price found with the independence coupling and the price found with the extreme coupling). This must be compared to the price that is obtained
when the cross-correlation matrix is taken of the form $\Sigma_{\rho} =
\left(%
\begin{smallmatrix}
\rho & \dots & \rho \\
\vdots &  & \vdots \\
\rho & \dots & \rho%
\end{smallmatrix}%
\right)$. As a matter of fact, the stress-test of the cross-correlation
fails, as the resulting correlation matrix $\left(%
\begin{smallmatrix}
Id & \Sigma_{\rho} \\
\Sigma_{\rho} & Id%
\end{smallmatrix}%
\right)$ is no longer definite positive when $\rho > \frac{1}{2\sqrt{3}}$
which is lower than 30\%. And even in the limit $\rho \to \frac{1}{2\sqrt{3}}
$, the price does not reach 0.075, and is still lower than the non-stressed
price.

\subsubsection{Indices of maximal correlation}

In order to better understand the link between the extreme coupling $\pi _{\hat{M},0}$ and the maximum correlation coupling (the one that corresponds to $M=Id$ in \eqref{maxCorrelPb}), we use a singular value decomposition of the affinity matrix $\hat{M}$ of the coupling $(X,Y)$. It writes $\hat{M}=USV^{\prime }$, with $U$ and $V$ two orthogonal matrices and $S$ a diagonal matrix with nonnegative entries. In particular,
\begin{equation*}
\mathbf{E}_{\pi _{\hat{M},0}}\Big((\sqrt{S}U^{\prime }X)^{\prime }(\sqrt{S}%
V^{\prime }Y)\Big)=\max_{\pi \in \Pi (P,Q)}\mathbf{E}_{\pi }\Big((\sqrt{S}%
U^{\prime }X)^{\prime }(\sqrt{S}V^{\prime }Y)\Big)
\end{equation*}%
In other words, if $(\tilde{X},\tilde{Y})=(\sqrt{S}U^{\prime }X,\sqrt{S}%
V^{\prime }Y)$, then this linear transform of $(X,Y)$ has maximum covariance
(under the distribution $\pi _{\hat{M},0})$.\bigskip

\noindent Thus if $\tilde{P}$ is the distribution of $\sqrt{S}U^{\prime }X$ with $X\sim P$, $\tilde{Q}$ is defined likewise from $Q$, and $\tilde{\pi}_{\hat{M},0}$ is
the distribution of $(\sqrt{S}U^{\prime }X,\sqrt{S}V^{\prime }Y)$ where $%
(X,Y)\sim \pi _{\hat{M},0}$, then $\mathbf{E}_{\tilde{\pi}_{\hat{M}%
,0}}(X^{\prime }Y)=\max_{\pi \in \Pi (\tilde{P},\tilde{Q})}\mathbf{E}_{\pi
}(X^{\prime }Y)$. Eventually, the singular value decomposition of the affinity matrix provides linear transforms of the marginals that makes the extreme coupling $\pi _{\hat{M},0}$ the maximum correlation coupling after a scaling of the marginals by these transforms. \newline
As an example, in the case of the 3 components described in the introduction of section \ref{introApplication}, this transform
writes
\begin{equation*}
\begin{array}{ccl}
\tilde{X} & = & \left(
\begin{smallmatrix}
- & 0.42 & X_{1} & +0.95 & X_{2} & -0.019 & X_{3} \\
- & 0.64 & X_{1} & -0.27 & X_{2} & +0.26 & X_{3} \\
& 0.11 & X_{1} & +0.06 & X_{2} & +0.35 & X_{3} \\
&  &  &  &  &  &
\end{smallmatrix}%
\right)  \\
\tilde{Y} & = & \left(
\begin{smallmatrix}
- & 0.30 & Y_{1} & +0.99 & Y_{2} & -0.13 & Y_{3} \\
- & 0.67 & Y_{1} & -0.16 & Y_{2} & +0..28 & Y_{3} \\
& 0.12 & Y_{1} & +0.08 & Y_{2} & +0.34 & Y_{3} \\
&  &  &  &  &  &
\end{smallmatrix}%
\right)
\end{array}%
\end{equation*}%
This result states that $\tilde{X}$ and $\tilde{Y}$ are most correlated to
one another under the distribution of the extreme coupling. These two
vectors are composed of portfolios involving the components of the original
index and can be viewed as new indices: we speak of \emph{indices of maximal
correlation}. When the strength of dependence is maximal ($T=0$), they
maximize the correlation $\mathbf{E}(\tilde{X}\tilde{Y})$ among all the
couplings with same marginals.\bk
This analysis can be seen as an analog in the case of fiexd multivariate marginals of the canonical correlation analysis, which consist, for two random vectors $%
X$ and $Y$, in finding vectors $a$ and $b$ such that the correlation between $a^{\prime}X$ and $b^{\prime} Y$ is maximal. In the multivariate setting, $\sqrt{S}U^{\prime }$ and $\sqrt{S}V^{\prime }$ are the analogue of the optimal $a$ and $b$. The technique described in this section has been introduced in the very different context of matching markets by \cite{DG} under the name \emph{saliency analysis}.

\section{Conclusion}

A recurring complaint in Applied Statistics is the \textquotedblleft curse
of dimensionality\textquotedblright: models that have a simple,
computationally tractable form in dimension one become very complex, both
computationally and conceptually in higher dimension. We show here that
convex analysis, along with the theory of Optimal Transport, can lead to
efficient solutions to the problem of extreme dependence. Building on a natural
geometric definition of extreme dependence, we have introduced an index of
dependence and used the latter to build stress-tests of dependence between
two sets of economic variables. This is particularly relevant in the case of
international finance, where the dependence between many economic variables
in two countries is of interest.

\section*{Acknowledgments}

The authors thank Rama Cont for a question which was the starting point of
this article and Guillaume Carlier and Alexander Sokol for helpful
conversation.

\newpage

\printbibliography

\newpage

\appendix

\section{Facts on conic orders}

\label{FactsConic}

In the space $\mathbf{M}_{I,J}(\mathbf{R})$, a \textit{basis} is a convex set $B$ with $0\notin \bar{B}$ (the closure of $B$). We assume that $B$ is a compact basis. 
 Let $K(B)$ be the \textit{dual cone} of the cone generated by $B$, $\mathbf{R}_+.B = \{\lambda.b,\, \lambda \geq 0,\, b \in B\}$, which means that: 
\[
	K(B) = \{\Sigma \in \mathbf{M}_{I,J}(\mathbf{R}) \vert \Sigma \cdot M \geq
0,\ M \in \mathbf{R}_+.B\}
\]
Its interior is 
\begin{equation*}
Int(K(B)) = \{\Sigma \in \mathbf{M}_{I,J}(\mathbf{R}%
)|\Sigma \cdot M>0,\ M\in \mathbf{R}_+.B \backslash \{0\}\}
\end{equation*}
It is important to note that in both definitions, $ \mathbf{R}_+.B$ and $\mathbf{R}_+.B\backslash \{0\}$ can be replaced by the basis $B$.
\bk
A \emph{strict partial order} is defined on $\mathbf{M}_{I,J}(\mathbf{R})$ by setting%
\begin{equation*}
M_1 \succ _{K} M_2\substack{\text{def} \\ \Leftrightarrow \\ \phantom{a}}M_1-M_2\in
K_{+}^{\ast }
\end{equation*}
If $S$ is a subset of $\mathbf{M}_{I,J}(\mathbf{R})$, a \emph{maximal}
element of $S$ for this order is a matrix $A\in S$ such that for all $B\in S$, $M_1 - M_2 \notin K_{+}^{\ast }$: $M_1$ can not be `strictly dominated' by any element in $S$.
\newline The choice of $\mathbf{M}_{I,J}(\mathbf{R})$ is arbitrary here and it could be replaced by any euclidean space.

\section{Proof of the results}

\label{ProofResults}

\subsection{Proof of Theorem \protect\ref{maxDep}}

\begin{proof}
As the covariance set is a closed convex set, a point $x\in \mathbf{M}_{I,J}(%
\mathbf{R})$ lies on its boundary if and only if there exists a non-zero $%
M\in \mathbf{M}_{I,J}(\mathbf{R}) \backslash \{0\}$ such that $M\cdot x$ is
maximal as a function of $x$. This translates the fact that there exists a
supporting hyperplane at $x$. Thus $\sigma _{\pi }$ is on the boundary of
the covariance set iff there exists $M\in \mathbf{M}_{I,J}(\mathbf{R})
\backslash \{0\}$ such that%
\begin{equation*}
M\cdot \sigma _{\pi }=\sup_{\tilde{\pi}\in \Pi (P,Q)}M\cdot \sigma _{\tilde{%
\pi}}
\end{equation*}%
(recall that $M\cdot \sigma _{\pi }=Tr\left( M^{\prime }\sigma
_{\pi }\right) $).
\newline Equivalence between (ii) and (iii) follows from a well-known result in
Optimal Transport theory, the Knott-Smith optimality criterion (see \cite%
{Villani:2003}, Th. 2.12). $\Box$
\end{proof}

\subsection{Proof of Theorem \protect\ref{maxUpperDep}}

Before we give the proof of the theorem, we state and prove a number of
auxiliary results which are of interest per se.
Let $B$ be a compact basis ; we have a crucial,
although technical, variational characterization of the maximality of $%
\sigma _{\pi}$ with respect to $\succ_{K(B)}$:

\begin{proposition}\textit{(Variational characterization of maximality)}
\label{variationalCharac}
\begin{equation*}
\text{$\sigma_{\pi}$ maximal iff } \inf_{M
\in B}\sup_{\tilde{\pi}\in \Pi(P,Q)}  (\sigma_{\tilde{\pi}} - \sigma_{\pi}) \cdot M = 0
\end{equation*}
\end{proposition}

In other words, a coupling is maximal whenever there exists $M\in B$ such
that $\sigma _{\pi }$ maximizes $\sigma _{\tilde{\pi}}\cdot M$.

\begin{proof}[Proof of proposition \protect\ref{variationalCharac}]
Note that for every $\pi \in \Pi(P,Q)$, the function%
\begin{equation*}
f: (\tilde{\pi},M)\in \Pi (P,Q)\times B \mapsto (\Sigma _{\tilde{\pi}}-\Sigma
_{\pi })\cdot M
\end{equation*}%
exhibits a saddlepoint $(\bar{\pi},\bar{S})$:%
\begin{equation}
\max_{\tilde{\pi}\in \Pi (P,Q)}\min_{M\in B}f(\tilde{\pi},M)=f(\bar{\pi},%
\bar{M})=\min_{M\in B}\max_{\tilde{\pi} \in \Pi (P,Q)}f(\tilde{\pi},M)
\label{minmax}
\end{equation}%
This is a consequence of a classical minmax theorem by \cite{Fan:51}: a
continuous function over a product of compact convex sets embedded in
normed linear spaces, which is linear in both arguments exhibits a
saddlepoint. Both $\Pi(P,Q)$ and $B$ are compact and convex. The compactness
of $B$ is an hypothesis, and it is a well-known fact that $\Pi (P,Q)$ is compact, see \cite{Villani:2003}. Moreover $f$ is linear in $M$ and $\tilde{\pi}$, and continuous in both arguments.
Finally, $\Pi (P,Q)$ can be embedded in the space of Radon measures over $%
\mathbf{R}^{I}\times \mathbf{R}^{J}$ endowed with the bounded Lipschitz
norm. We refer to Villani \textit{op. cit.} chapter 7. for more details on
this point: the important thing is that $\Pi (P,Q)$ is a compact subset (for this norm) of this space.\bk
Back to the proof of the theorem. If $\sigma _{\pi }$ is maximal, then for
all $\sigma _{\tilde{\pi}}$ one has $\sigma _{\tilde{\pi}}-\sigma _{\pi
}\notin Int(K(B))$, which means that for some $M\in B$, $(\sigma _{%
\tilde{\pi}}-\sigma _{\pi })\cdot M\leq 0$, hence
\begin{equation*}
\sup_{\tilde{\pi}\in\Pi (P,Q)}\inf_{M \in B}(\sigma _{\tilde{\pi}}-\sigma _{\pi })\cdot M\leq 0
\end{equation*}%
And therefore the above quantity is necessarily zero, because one may choose $\tilde{\pi} = \pi$. Thanks to the compactness of $B$ and $\Pi(P,Q)$, the minmax theorem applies and yield that the infimum of the supremum is zero.
\newline On the contrary, if $\sigma _{\pi}$ is not maximal then there exists some coupling $\tilde{\pi}$ such that $\sigma_{\tilde{\pi}}-\sigma_{\pi }\in Int(K(B))$. Thus, for all $M \in B$, $\sup_{\tilde{\pi \in \Pi(p,q)}}\sigma_{\tilde{\pi}}-\sigma_{\pi } \cdot M > 0$, and thanks to the compactness of $B$, 
\[
	\inf_{M \in B}\sup_{\tilde{\pi} \in \Pi(p,q)}\sigma_{\tilde{\pi}}-\sigma_{\pi } \cdot M > 0
\]

\end{proof}
As a consequence, we are now ready to prove theorem \ref{maxUpperDep}.

\begin{proof}[Proof of theorem \protect\ref{maxUpperDep}]$\phantom{a}$\newline
$(ii) \Rightarrow (i)$: 
If for some $M \in B$, a coupling $\pi$ satisfies
\begin{equation*}
\mathbf{E}_{\pi }(X\cdot MY)=\sup_{\tilde{\pi}\in \Pi (P,Q)}\mathbf{E}_{%
\tilde{\pi}}(X\cdot MY)
\end{equation*}
then $\sup_{\tilde{\pi}\in \Pi (P,Q)}(\sigma_{\tilde{\pi}} - \sigma_{\pi})\cdot M = 0$ and so $\inf_{M \in B}\sup_{\tilde{\pi}\in \Pi (P,Q)}(\sigma_{\tilde{\pi}} - \sigma_{\pi})\cdot M \leq 0$. But this is an infimum of quantities that are greater than zero, and eventually the `inf sup' is zero. \bk
$(i) \Rightarrow (ii)$: if $\sigma_{\pi}$ is maximal, then proposition \ref{variationalCharac} entails $\inf_{M \in B}\sup_{\tilde{\pi} \in \Pi(p,q)}\sigma_{\tilde{\pi}}-\sigma_{\pi } \cdot M = 0$. Due to the compactness of $B$, there exists a matrix $M \in B$, such that the supremum is zero, which concludes the proof of this implication.
$\Box$
\end{proof}

\section{More details on problem \eqref{minProblem}}
\label{detailsMinProblem}
The objective function of the problem \eqref{minProblem} is convex in $M$, because it is the sum of: a linear function of $M$,$-\sigma_{\hat{\pi}} \cdot M$ ; and of $W(M,1)$, which is convex in $M$ as the supremum over $\pi \in \Pi(p,q)$ of linear functions in $M$, namely $\esp_{\pi}(X'MY) = \sigma_{\pi} \cdot M$.
\newline Moreover, assuming that the entropy of the coupling $\hat{\pi}$ is finite, then $W(M,1) \geq  \sigma_{\hat{\pi}} \cdot M + \textrm{Ent}(\hat{\pi})$. Thus $W(M,1) - \sigma_{\hat{\pi}} \cdot M \geq \textrm{Ent}(\hat{\pi}) > -\infty$. A convex function which is bounded below admits a global minimizer.
\newline Moreover, the objective function is differentiable as $W(M,1)$ is differentiable and $\nabla_M W(M,1) = \sigma_{\pi(M,1)}$. This is a consequence of a property of subdifferentials, see e.g. \cite{valadier}. A global minimizer is necessarily a critical point, proving that the solution $M$ of problem \eqref{minProblem} satisfies $\sigma_{\pi(M,1)} = \sigma_{\hat{\pi}}$.
\newline Nevertheless, depending on the marginal distributions, this minimizer might not be unique: for instance if $P$ is the law of a vector $(X_1,0)$ where $X_1$ is a Gaussian random variable. The second row of the matrix $M$ does not matter here, and this prevents problem \eqref{minProblem}  from having a unique solution.
\end{document}